\documentclass[sigplan,nonacm]{acmart}
\makeatletter
\AtBeginDocument{%
  \fancyhead[LO]{}%
  \fancyhead[RE]{}%
}
\makeatother
\usepackage{amsmath}
\usepackage{xspace}
\usepackage{placeins}
\usepackage{stfloats}
\newcommand{\system}{LearnSched\xspace}
\newcommand{\pos}[1]{\left(#1\right)_{+}}

\title{Serve Now or Improve Later? Scheduling Self-Evolution in Online Agent Systems}

\author{Yangbo Wei}
\affiliation{%
  \institution{Shanghai Jiao Tong University}
  \city{Shanghai}
  \country{China}}
\affiliation{%
  \institution{Eastern Institute of Technology}
  \city{Ningbo}
  \country{China}}

\author{Junhong Qian}
\affiliation{%
  \institution{Shanghai Jiao Tong University}
  \city{Shanghai}
  \country{China}}

\author{Zhen Huang}
\affiliation{%
  \institution{University of Science and Technology of China}
  \city{Hefei}
  \country{China}}

\author{Zhenyu Su}
\affiliation{%
  \institution{Shanghai Jiao Tong University}
  \city{Shanghai}
  \country{China}}
\affiliation{%
  \institution{Eastern Institute of Technology}
  \city{Ningbo}
  \country{China}}

\author{Qifan Wang}
\affiliation{%
  \institution{Shanghai Jiao Tong University}
  \city{Shanghai}
  \country{China}}
\affiliation{%
  \institution{Eastern Institute of Technology}
  \city{Ningbo}
  \country{China}}

\author{Shaoqiang Lu}
\affiliation{%
  \institution{Shanghai Jiao Tong University}
  \city{Shanghai}
  \country{China}}
\affiliation{%
  \institution{Eastern Institute of Technology}
  \city{Ningbo}
  \country{China}}

\author{Rumin Zhang}
\affiliation{%
  \department{Ningbo Institute of Digital Twin}
  \institution{Eastern Institute of Technology}
  \city{Ningbo}
  \country{China}}

\author{Chen Wu}
\affiliation{%
  \department{Ningbo Institute of Digital Twin}
  \institution{Eastern Institute of Technology}
  \city{Ningbo}
  \country{China}}

\author{Lei He}
\affiliation{%
  \institution{Eastern Institute of Technology}
  \city{Ningbo}
  \country{China}}

\makeatletter
\renewcommand{\@mkauthors}{%
  \global\setbox\mktitle@bx=\vbox{%
  \hsize=\textwidth
  \unvbox\mktitle@bx
  \par\medskip
  \centering
  {\normalfont\normalsize
    \noindent
    \mbox{Yangbo Wei\textsuperscript{1,2}}\quad
    \mbox{Junhong Qian\textsuperscript{1}}\quad
    \mbox{Zhen Huang\textsuperscript{4}}\quad
    \mbox{Zhenyu Su\textsuperscript{1,2}}\quad
    \mbox{Qifan Wang\textsuperscript{1,2}}\quad
    \mbox{Shaoqiang Lu\textsuperscript{1,2}}\quad
    \mbox{Rumin Zhang\textsuperscript{3}}\quad
    \mbox{Chen Wu\textsuperscript{3,*}}\quad
    \mbox{Lei He\textsuperscript{2,*}}\par}
  \vspace{4pt}
  {\normalfont\small
    \noindent\textsuperscript{1}Shanghai Jiao Tong University\par
    \noindent\textsuperscript{2}Eastern Institute of Technology\par
    \noindent\textsuperscript{3}Ningbo Institute of Digital Twin, Eastern Institute of Technology\par
    \noindent\textsuperscript{4}University of Science and Technology of China\par}

  }%
}
\makeatother

\begin{document}
\begin{abstract}
Online agents can improve future service by constructing reusable tools, guidance, or model states, but this work competes with current requests for the same GPUs.
Exploiting idle compute for self-evolution faces a fundamental systems constraint: benefits arrive only after an artifact is published and used, while pausing evolution leaves service capacity waiting for memory release and runtime recovery.
An investment worth completing may therefore be worth postponing.
We present \system, a state-aware scheduler that brings the reuse window of a capability and the timely return of compute into a common investment model.
\system incorporates candidate progress, checkpoint overhead, and the current recovery path into action values. Under the same information and capacity constraints, it uses one-step counterfactual rollout to choose progress, checkpointing, or waiting relative to a fully costed window policy.
We characterize handoff costs through independent A100 component measurements and evaluate action selection in 1920 paired finite-model scenarios.
When cold recovery is expensive, waiting for longer execution windows can improve net value by avoiding handoffs; in evaluated warm-recovery scenarios, the strong window policy already achieves the same value.
Retaining recoverable state can shorten capacity return and reduce the need for complex evolution scheduling.
\end{abstract}
\maketitle

\section{Introduction}
An online coding agent repeatedly fails on the same class of requests. It can continue invoking a model for each request, or use prior trajectories to construct a reusable tool or guidance, validate it, and publish it for later use. Persistent demand can repay this investment through fewer model calls and better answers; short-lived demand may leave even a qualified artifact unused. When service and evolution share GPUs, construction also consumes compute and memory needed by current requests. The system must decide: \emph{should the next unit of compute serve now or improve later?}

\begin{figure}[t]
\centering
\includegraphics[width=\columnwidth]{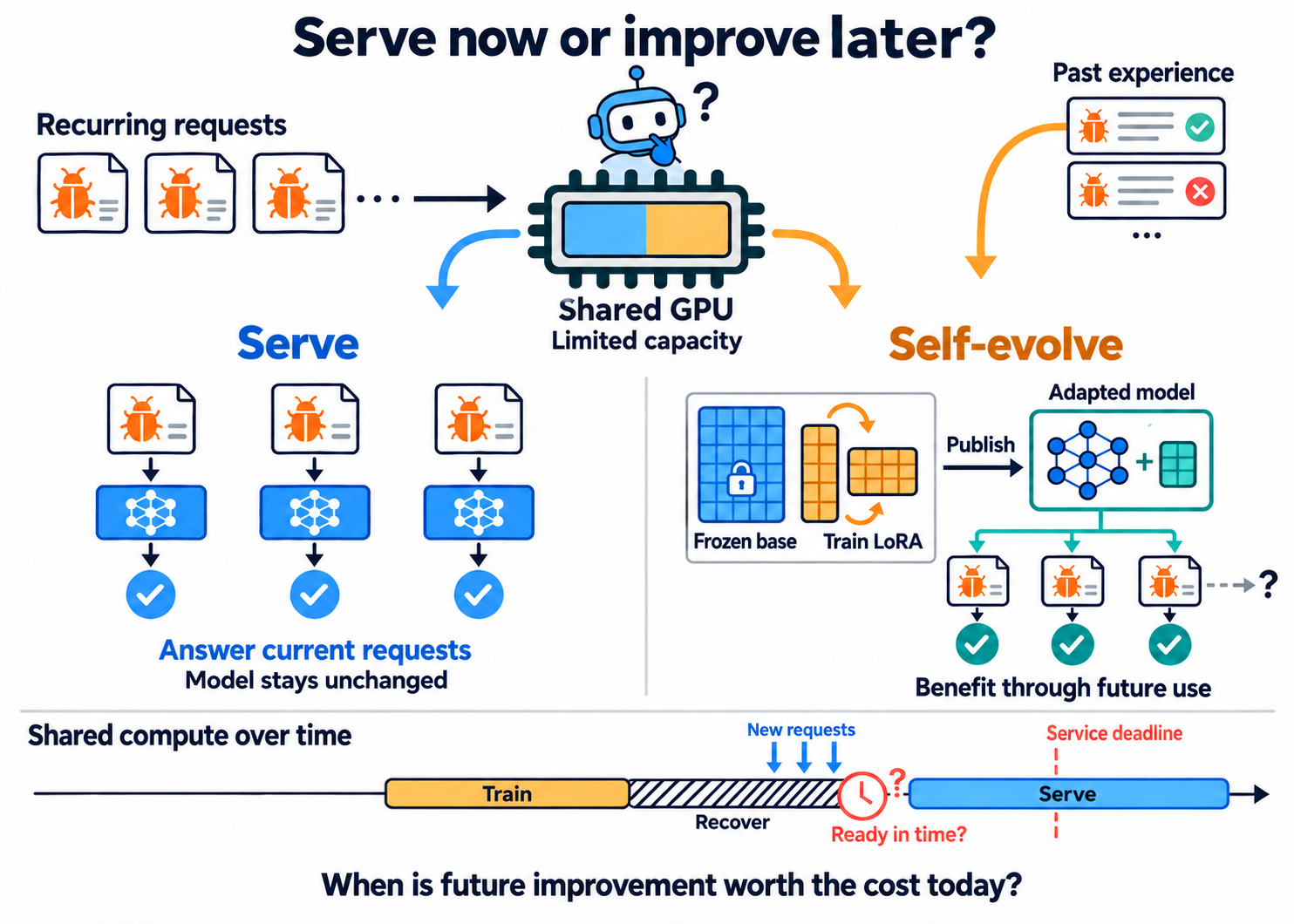}
\caption{Serving and self-evolution compete for shared GPU capacity. LoRA adaptation illustrates the tradeoff: invest in future capability while returning compute in time to meet current request deadlines.}
\Description{A shared GPU branches into serving current requests with an unchanged model and self-evolution using past experience to train LoRA adapters while freezing the base model. The adapted model serves future requests. A timeline shows training, recovery, and serving, with new requests arriving during recovery and a later service deadline.}
\label{fig:intro1}
\vspace{-8pt}
\end{figure}

Figure~\ref{fig:intro1} illustrates this choice through LoRA adaptation: improving the model incurs training and recovery costs before future requests can benefit.

This decision couples two timescales. Foreground requests consume resources under latency deadlines, while capability construction incurs costs now for benefits spread across future requests. A useful artifact can reduce service work, but only after qualification, publication, and actual use. Its benefit cannot compensate for requests that have already missed their deadlines. The scheduling problem therefore involves both the value of the investment and the service capacity available along its execution path.

Running evolution during idle periods and pausing it at the next burst is a natural starting point. Yet an idle interval has two distinct boundaries: the latest start that leaves enough reuse to amortize construction, and the latest return that restores capacity before service needs it. A pause command may leave device memory occupied; releasing memory may still leave a serving runtime to reconstruct. \emph{Stopping evolution, reclaiming memory, and making capacity ready to serve are distinct system events.} Treating the end of background computation as the return of service capacity overstates the useful part of an execution window.

Runtime state connects these boundaries. Termination discards unsaved progress, checkpointing consumes time, and retaining service state trades memory for faster recovery. Consider a short idle window followed by one long enough to complete construction. Immediate progress can publish the artifact slightly earlier but incur an extra checkpoint and cold recovery. Waiting pays when the avoided handoff costs exceed the benefits lost to delay. Cheap warm recovery can reverse this choice without changing the candidate or future demand. Thus, \emph{an investment worth completing need not be worth starting now.}

Expected future benefits have long guided adaptive compilation~\cite{jikesrvm}, inference and retraining allocation~\cite{ekya,recl}, and agent learning during idle windows~\cite{metaclaw}. Evolution control planes also incorporate cost signals~\cite{evolutionplane}. We study the interaction between this investment perspective and the runtime mechanisms that preserve progress and restore service. With the same candidate, information, and execution mechanisms, when do durable progress and recovery costs change the best execution plan? Answering this question requires separating gains from better accounting, faster recovery, and different scheduling actions.

\begin{figure*}[!tbp]
\centering
\includegraphics[width=\textwidth]{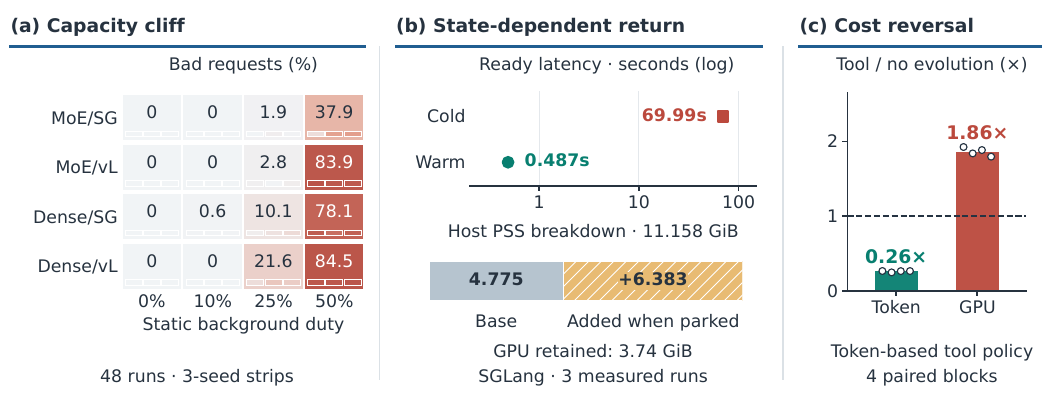}
\caption{Three investment constraints: (a) bad-request rates across four systems and four shares (three repeats; SG/vL: SGLang/vLLM); (b) cold/warm readiness means and min--max, with host PSS and GPU residency separate; (c) token and GPU cost ratios to no evolution, across four paired blocks. GPU activity excludes CPU costs.}
\Description{A three-panel motivation figure showing capacity contention, state-dependent recovery latency and resident memory, and the reversal between token and device costs.}
\label{fig:motivation}
\vspace{-5pt}
\end{figure*}

We present \system, a state-aware scheduler that compares remaining benefits and execution costs under a capacity-return constraint. Its model connects the useful lifetime of a capability to completed and durable progress, current service state, and the recovery path. These quantities determine whether to advance, checkpoint and return, discard unsaved work, or wait. Execution components expose work pressure, publication and use, and capacity readiness to account for costs and benefits as they occur.

\system uses classical one-step rollout~\cite{rollout1997} relative to Costed Window, a reference that fully prices a greedy completion plan. Both receive the same windows and recovery mechanisms. Comparing alternative continuations allows \system to skip a costly short window or change execution segments when this improves value; retaining the reference action guarantees no loss in the accurate finite model. This formulation makes the added value of scheduling measurable against an already cost-aware reference.

Our evaluation separates independent A100 component measurements and capability ledgers from paired finite-model scheduling. Across 1920 scenarios, full cost accounting explains most of the benefit. Waiting adds value when cold recovery is expensive and longer windows lie ahead; all evaluated warm-recovery scenarios tie with Costed Window. The resulting systems implication is that state retention changes both return latency and the amount of value available to more complex planning. Runtime memory and investment policies thus require joint evaluation.

Our contributions are:
\begin{itemize}
\item A model connecting capability break-even conditions, durable progress, and state-dependent capacity return to explain when waiting improves net value.
\item A scheduler that compares progress and waiting under common constraints, with finite-model policy improvement and an explicit scoring-error condition.
\item Component measurements, capability ledgers, and paired evaluation isolating accounting and scheduling gains and their dependence on recovery state.
\end{itemize}

\section{Why Timing and State Matter}
\label{sec:motivation}
Figure~\ref{fig:motivation} examines three constraints on using shared GPUs for evolution. Positive expected return and available idle time alone do not determine the execution plan.

\paragraph{Capacity must be available when requests need it}
Raising the static background share from 10\% to 25\% increases bad requests across all four systems; at 50\%, mean rates reach 37.88--84.48\% (Figure~\ref{fig:motivation}(a)). An average resource budget hides the periods when foreground demand and background execution overlap. A profitable capability can therefore degrade service before its benefits arrive. The scheduler needs to control the timing of resource use as well as its volume: capacity borrowed under low load must be returned before the next service deadline, and future work savings cannot be credited before the artifact is used.

\paragraph{Runtime state changes the usable window}
Cold service readiness takes 69.99 s on average, whereas same-process warm recovery takes 0.487 s (Figure~\ref{fig:motivation}(b)). These paths leave substantially different amounts of an idle window for useful evolution work. Segmentation compounds the difference: each return can incur recovery, and interrupted construction may require checkpointing or recomputation. Two plans that complete the same amount of work can consequently have different total costs and publication times.
Warm recovery also raises host proportional set size (PSS) from 4.77 to 11.16 GiB and retains 3.74 GiB on the device. Keeping state buys faster return by occupying memory that other work could use. Plan comparisons must jointly price recovery latency, durable progress, and residency; a fixed pause overhead cannot capture their tradeoff.

\paragraph{The constrained resource determines the cost of reuse}
Under sustained tool demand, token-based amortization reduces token cost to 0.2624 times that of no evolution but raises GPU active time to 1.8623 times (Figure~\ref{fig:motivation}(c)). Construction and the service calls it replaces have different device costs per token. Counting avoided calls or tokens can therefore reverse the resource-cost ranking. This observation is a cost comparison; quality gains and CPU execution still matter to application utility. Likewise, publishing more artifacts creates value only when subsequent requests use them. Investment accounting must connect construction to realized use and price both in the constrained resource.

These observations motivate a joint decision over a candidate, an execution interval, and a return path. Starting immediately captures earlier benefits, while waiting can avoid checkpointing and repeated recovery. In the paired finite-model example of Section~\ref{sec:state_eval}, delaying publication by 4.952 s saves 70.104 s-eq. in execution and handoff costs under the same candidate and calendar. The relevant comparison is between complete feasible plans, including their return paths and the benefits forgone while waiting.

\section{Delayed Returns and Switching Costs}
\label{sec:model}
We model one candidate capability on shared devices. Actions change progress, publication time, and recovery state; request deadlines constrain capacity return.

\subsection{State and Continuation Value}
At time $t$, the state is
\begin{equation}
 s_t=(m_t,p_t,\bar p_t,z_t,q_t),\qquad 0\leq\bar p_t\leq p_t.
 \label{eq:state}
\end{equation}
Here, $m_t$ is service or evolution mode, $p_t$ is completed progress, $\bar p_t$ is durable progress, $z_t$ identifies resident service state and its recovery path, and $q_t$ is foreground backlog. Termination loses $p_t-\bar p_t$; checkpointing preserves it. Previously paid compute is a sunk cost. Information $\mathcal I_t$ supports estimates of future demand, costs, and transitions.

Actions wait, advance work, checkpoint and return, or discard unsaved work and return. Action $a$ takes $\tau(s,a)$; $\mathcal A_{\rm ret}(s,\mathcal I_t)$ contains actions meeting the capacity-return constraint. With benefit $r$ and cost $c$ in a common unit,
\begin{equation}
 \begin{split}
 V_t^*(s)=\max_{a\in\mathcal A_{\rm ret}}
 \{r(s,a)-c(s,a)\\
 {}+\mathbb E[V_{t+\tau(s,a)}^*(s')\mid s,a,\mathcal I_t]\}.
 \end{split}
 \label{eq:bellman}
\end{equation}
Benefits arise after publication and use. Cost $c$ charges only the current action's execution, checkpointing, recovery, and residency; later costs enter continuation value. In the deterministic single-candidate instance, with no execution choices after publication at $t_p$, the remaining benefit stream can be settled as terminal value $\Phi(t_p)=\int_{t_p}^{H}g(v)\,dv$. The publication action still pays its return cost.

\subsection{Effective Reuse and Break-Even}
Fix a continuous path that qualifies and publishes after $\tau_e$, at remaining cost $C_e>0$. Demand ends at $H$; matching requests arrive at rate $\lambda$, fraction $a$ use the artifact, and each use has expected net value $u$ relative to base service. Then
\begin{equation}
 V(t)=M(t)u-C_e,\qquad M(t)=\lambda a\pos{H-t-\tau_e}.
 \label{eq:value}
\end{equation}
For $\lambda au>0$, positive direct return requires
\begin{equation}
 t<t_e^*=H-\tau_e-\frac{C_e}{\lambda au}.
 \label{eq:expiry}
\end{equation}
The deadline reserves time for both completion and amortization. Nonpositive unit value, absent reuse, or a start at or beyond this deadline prevents positive direct return along the fixed path. Changing segments or recovery paths requires evaluating their full continuation costs.

For uncertain qualification, publication, and use, let $Y_i$ indicate their joint occurrence for request $i$, with use value $u_i$:
\begin{equation}
 \mathbb E[V\mid\mathcal I_t]=\sum_i\Pr(Y_i=1\mid\mathcal I_t)
 \mathbb E[u_i\mid Y_i=1,\mathcal I_t]-\mathbb E[C_e\mid\mathcal I_t].
 \label{eq:conditional}
\end{equation}
The joint event retains correlations between completion time and remaining use opportunities. The cost term includes successful and failed construction paths. New observations update this estimate at stage boundaries.

\subsection{Request Deadlines and Lendable Capacity}
Consider an initially empty, work-conserving FIFO fluid queue with one resource~\cite{netcal}. Request $i$ arrives at $a_i$, needs work $w_i>0$, and has deadline $d_i\geq a_i$; service rate is $c_{\rm cap}(v)$. Timely completion requires and is guaranteed by $\sum_{j=k}^{i}w_j\leq\int_{a_k}^{d_i}c_{\rm cap}(v)\,dv$ for every $k\leq i$.
Relative to base service on the same arrivals, define $S_{ki}$ as baseline capacity minus work over this interval, $R_{ki}$ as work actually avoided through capability use, and $B_{ki}$ as capacity lost to contention and handoff. Equivalently,
\begin{equation}
 B_{ki}\leq S_{ki}+R_{ki},\qquad\forall k\leq i.
 \label{eq:capacity}
\end{equation}
Future benefits add no current capacity; use that adds work makes $R_{ki}$ negative. Stopping a candidate reduces $B_{ki}$ only when capacity returns to service. Our component approximates pressure through in-flight work; batching and CPU queues need further modeling, and unknown arrivals preclude a hard deadline guarantee.

\subsection{State-Dependent Return Deadlines}
If additional serving capacity must be ready at $t_b$, a serial return path takes
\begin{equation}
 \tau_r=\tau_{\mathrm{act}}+\tau_{\mathrm{release}}+\tau_{\mathrm{ready}}.
 \label{eq:return}
\end{equation}
The terms denote command actuation, resource release, and runtime readiness. Overlapping steps follow the critical path; residency can avoid runtime reconstruction. With deterministic delay,
\begin{equation}
 t_{\mathrm{reclaim}}\leq t_b-\tau_r.
 \label{eq:return_deadline}
\end{equation}
Checkpointing further reduces execution time; discarding work makes later execution resume from durable progress. Recovery state therefore changes both feasible actions and their costs. Under uncertain bursts, early return trades progress against avoidable service loss, as the two-action model in Appendix~\ref{app:risk} shows.

\begin{figure*}[!tb]
\centering
\includegraphics[width=\textwidth]{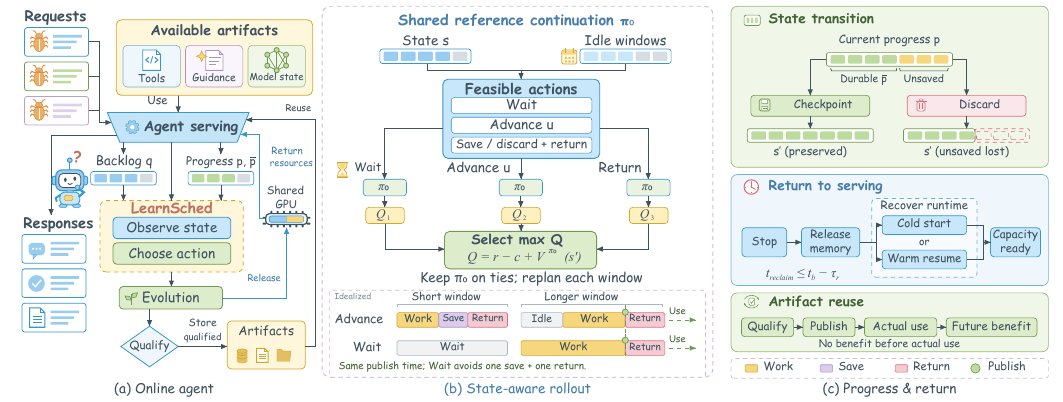}
\caption{\system overview. (a) Serving and evolution share GPU capacity; the scheduler observes backlog and progress, controls evolution, and returns resources to serving. Qualified artifacts are retained for reuse. (b) Safe actions are scored under the same reference continuation $\pi_0$. The idealized timelines align publication times: waiting avoids an extra checkpoint and return. (c) Checkpointing preserves progress; discarding loses only unsaved work. Capacity becomes ready after memory release and cold or warm recovery; benefits require actual artifact use.}
\Description{Three panels show the online agent control loop, state-aware rollout, and progress and return mechanisms. Serving consumes requests and reusable artifacts; LearnSched observes backlog and progress to control evolution on shared GPUs. Feasible actions follow a common reference continuation and are compared by net continuation value. Two idealized timelines publish at the same time, while waiting avoids one checkpoint and one return. Checkpointing retains all completed progress, whereas discarding resets progress to its durable prefix. Memory release precedes either cold start or warm resume and capacity readiness. Artifact benefits follow qualification, publication, and actual use.}
\label{fig:overview}
\end{figure*}

\subsection{Common Pricing}
If a policy gains $\Delta Q_{\mathrm{timely}}$ correct answers on time at additional cost $\Delta C_m$ in resource unit $m$, its incremental utility is
\begin{equation}
 \Delta U(W)=W\Delta Q_{\mathrm{timely}}-\Delta C_m
 \label{eq:utility}
\end{equation}
Here, $W$ is the application value of a timely correct answer, in $m$/answer. Tokens, GPU active time, and host memory are measured separately and combined only through explicit prices. For background steps uncalibrated to capability benefits, we report the progress--bad-request frontier directly. Model costs use equivalent evolution seconds (s-eq.).

\section{\system: State-Aware Investment Scheduling}
\label{sec:design}
\system compares progress, checkpointing, and waiting through their continuation values. Execution components estimate reuse, bound in-flight work, and manage handoff.
Figure~\ref{fig:overview} connects the online control loop, rollout decision, and state-dependent return path.

\subsection{Window Actions and the Reference}
Known idle windows $[\ell_k,d_k]$ define decision stages with state $(k,p,\bar p,m;z)$. A macro-action advances $u$ quanta of duration $\delta$ and optionally checkpoints incomplete work, subject to
\begin{equation}
 \tau_{\rm load}+u\delta+
 \mathbf 1_{\rm save}\tau_{\rm ckpt}+\tau_{\rm return}
 \leq d_k-\ell_k.
 \label{eq:macro_safe}
\end{equation}
Checkpointing and recovery consume exact wall-clock time. The next window starts in service mode; unsaved work is discarded, and published artifacts leave construction. Waiting in service mode adds no recovery cost.

All policies share windows, benefits, and execution mechanisms. Window advances as far as feasible and checkpoints. Costed Window ($\pi_0$) prices all remaining benefits and costs of that greedy completion plan against abandoning the candidate and returning capacity. It executes if completion has greater value; otherwise, it waits and reevaluates. Static ROI omits handoff costs from valuation only.

\subsection{Rollout and Policy Improvement}
\system scores each safe current action by following the reference thereafter~\cite{rollout1997}:
\begin{equation}
\begin{aligned}
 Q^{\pi_0}(s,a)=r(s,a)-c(s,a)+V^{\pi_0}(s'),
 \\
 \pi_R(s)\in\arg\max_a Q^{\pi_0}(s,a).
\end{aligned}
 \label{eq:rollout}
\end{equation}
The candidate set includes $\pi_0(s)$, retains it on ties, and is recomputed at each window.

\paragraph{Proposition 1: Policy improvement}
With finitely many stages, accurate transitions, costs, and reference values, and a safe $\pi_0(s)$, every reachable state satisfies
\begin{equation}
 V^{\pi_R}(s)\geq V^{\pi_0}(s).
 \label{eq:improvement}
\end{equation}
Induction on remaining stages gives
$V^{\pi_R}(s)\geq Q^{\pi_0}(s,\pi_R)\geq Q^{\pi_0}(s,\pi_0)=V^{\pi_0}(s)$,
starting from equal terminal values.
If $|\widehat Q-Q^{\pi_0}|\leq\epsilon$ uniformly, accepting only changes with $\widehat Q(s,a)>\widehat Q(s,\pi_0)+2\epsilon$ also ensures nondecreasing true scores. The evaluated model uses accurate scores; an online guarantee requires a validated error bound.

\subsection{Waiting Value and Computation}
Let immediate and delayed plans publish the same artifact at $t_1$ and $t_2\geq t_1$, before $H$, with remaining costs $C_1,C_2$ and fixed subsequent benefit rate $g>0$:
\begin{equation}
 V_{\rm wait}-V_{\rm now}=(C_1-C_2)-g(t_2-t_1).
 \label{eq:wait_value}
\end{equation}
Waiting pays when avoided handoff and checkpoint costs exceed lost benefits. With one sufficiently long window, free switching and checkpointing, and no information updates, the choice reduces to comparing $g(H-t-\tau_e)-C_e$ with zero.
Figure~\ref{fig:overview}(b) illustrates the equal-publication-time case, in which waiting saves one checkpoint and one return.
The implementation enumerates $O(J)$ progress budgets over at most $K$ remaining windows and caches continuation values for equal durable progress. Exact DP over all reachable states and actions gives the model optimum.

\subsection{Observation and Admission}
Candidates record task class, remaining stage and cost, completion time, and reuse opportunities. Tool admission uses historical statistics and current arrivals; guidance uses class shares and quality proxies calibrated on independent history. Hidden target tests and future task identifiers do not enter estimates.
The prototypes reevaluate value while synthesis is pending and check windows and capacity thereafter. Issued calls are charged through completion; full ROI optimization at each validation stage is not implemented.

Admission estimates work as $w=L_{\max}+0.0625P$, using prompt length $P$ and maximum output length $L_{\max}$. A stage runs only if it fits with in-flight work within budgets of 1536 for tools and 1024 for guidance. This maximum-output reservation can be conservative. A separate contention controller uses arrival-aware additive increase and multiplicative decrease (AIMD) to adapt the background share.

\subsection{Handoff and Publication Accounting}
Pollable waits insert control checks into long sleeps. Whole-GPU handoff checkpoints LoRA, terminates training, confirms device-memory release, and starts a serving replica. Training resumes from the final durable file published after termination, avoiding a checkpoint-publication race. An alternative retains the serving process and restores its device state on return, trading resident memory for faster recovery.

Tools publish \texttt{solve(x)} after passing eight public inputs; each configuration starts with an empty library. Guidance generates base and additional answers, preferring a base answer that passes public tests. Each request freezes its visible artifacts before dispatch. Model-call accounting charges actual service, extra generation, construction, and validation:
\begin{equation}
 C_{\mathrm{all}}=C_{\mathrm{base}}+C_{\mathrm{extra}}+
 C_{\mathrm{build}}+C_{\mathrm{validate}}.
 \label{eq:ledger}
\end{equation}
Failed builds, unpublished candidates, and unused calls are charged; absent base calls are not. Unknown usage remains missing. CPU execution is timed separately; the handoff ledger records durable progress, unsaved work, readiness, and device and host memory.

\section{Experimental Setup}
\label{sec:setup}
Component experiments measure contention, execution costs, and effective reuse. Comparisons use the same capabilities and resources, except Cal.*, which receives the future arrival calendar. Additional ledgers and sensitivity details appear in the appendices.

\subsection{Hardware and Workloads}
Contention experiments cross Granite-3B-A0.8B (MoE) and Qwen2.5-Coder-7B (Dense) with vLLM and SGLang on one A100-80GB. Arrivals follow 4$\rightarrow$40$\rightarrow$4 requests/s for 10 s per phase, with at most 64 output tokens. SLOs are 1.2, 0.62, 1.8, and 1.2 s for MoE on SGLang and vLLM, then Dense on SGLang and vLLM. Bad requests include deadline misses from scheduled arrival and failures: the 60 runs contain 28,805 requests, including two failures. Background steps count synthetic GPU work without capability calibration.

Tool experiments use Qwen3.5-9B on SGLang, a 512-token output limit, zero temperature, and no radix cache. In each 24 s window, sustained demand has 48 requests per block, stopped demand retains the same 12-request prefix, and sparse demand has requests at seconds 6 and 18. Configurations rebuild tools, isolate public tests from hidden evaluation, and run CPU paths in isolated workers.

For guidance, we dispatch 114 known tasks from LiveCodeBench~\cite{livecodebench} under low--high--low arrivals. Five policies and five blocks each give 5700 requests.

Whole-GPU handoff uses two A100s, Granite 3B, and rank-8 LoRA. One GPU serves; the other alternates between training and a replica. The vLLM trace has 100 s of low load, a 60 s burst at 120 requests/s, and 30 s of low load, with at most 128 output tokens and a two-second SLO. Reactive handoff starts at burst onset; advance-notice handoff starts 90 s earlier, with experiment-specified notice. SGLang recovery is measured independently.

\subsection{Software}
Inference uses vLLM and SGLang. Python control uses asyncio and aiohttp for asynchronous dispatch and cost/latency recording; tool validation and execution use subprocesses with resource limits. The finite-state simulator implements Window, Costed Window, rollout, and exact DP under common transition and cost models.

\section{Evaluation}
\label{sec:evaluation}
We evaluate plan selection, resource admission, capacity return, and investment payback. Each experiment compares policies under its stated information and cost model.

\subsection{Value of State-Aware Scheduling}
\label{sec:state_eval}
All policies share complete window information, transitions, costs, and return constraints; exact DP gives the optimum in the finite macro-action model.
\begin{figure*}[!tbp]
\centering
\includegraphics[width=\textwidth]{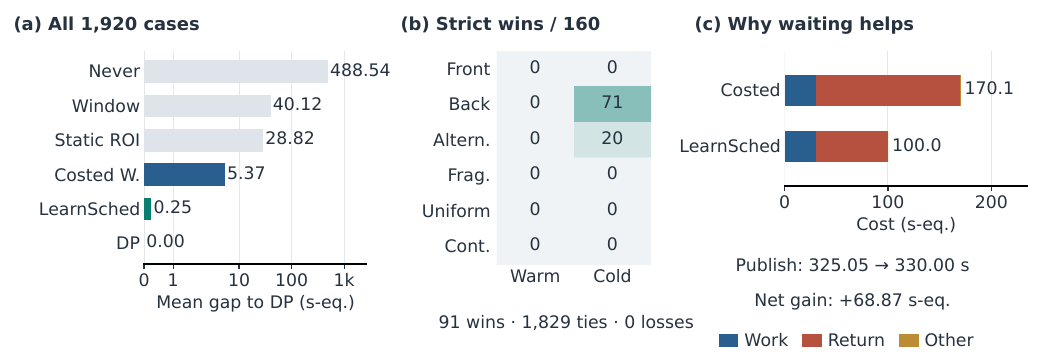}
\caption{Paired finite-model results. (a) Mean DP gaps over 1920 scenarios (symlog). (b) Strict wins over Costed Window per 160 scenarios. (c) S0348, the first improved fresh case: waiting avoids one cold recovery. Other: save, load, residency.}
\Description{Model optimality gaps for six policies, improvements across recovery states and six calendars, and a paired example where waiting saves recovery costs.}
\label{fig:state_policy}
\end{figure*}

\paragraph{Scenarios and costs}
The frozen grid crosses six calendars (80--480 s windows separated by 20 s), four work sizes (30/60/120/180 s), five benefit rates (0.1/0.25/0.5/1/2 s-eq./s), two benefit tails (120/480 s), cold/warm recovery, and four initial states: fresh, half-complete and unsaved, saved and active, or saved in service mode. This gives 1920 scenarios with a 5 s work quantum; Appendix~\ref{app:model_grid} specifies the calendars.
Cold/warm recovery uses three-run SGLang means of 69.988/0.487 s. Independent LoRA measurements over 20 runs give save/load costs of 68.075/47.748 ms for about 13.24 MB of state. Execution and recovery cost one s-eq. per second; warm recovery adds 6.383 GiB of host PSS priced at 0.01 s-eq. per GiB-second. These separate measurements are combined in the model; memory pricing is assumed, unmeasured entry/release delays are zero, and resident-memory throughput effects are not fitted.

\paragraph{Aggregate value and distribution}
Figure~\ref{fig:state_policy}(a) reports mean DP gaps of 40.122 for Window, 28.819 for static ROI, 5.369 for Costed Window, and 0.246 s-eq. for rollout. Rollout raises mean net value from 465.547 to 470.670, gaining 5.123 s-eq. (1.10\%): full cost accounting explains most of the benefit.
It improves 91 scenarios, ties in 1829, and never regresses. All improvements involve cold recovery: 71 back-loaded and 20 alternating calendars (Figure~\ref{fig:state_policy}(b)); all 960 warm scenarios tie. Rollout matches DP in 1907 scenarios, but the remaining 13 have a maximum gap of 124.836. Every policy meets return deadlines because they share the same action constraint. Never can incur negative value from returning initially active states.
At a 1 s quantum in 480 fresh-candidate scenarios, rollout improves 27 and ties in 453; its mean DP gap is zero, versus 5.869 for Costed Window.

\begin{figure*}[!tbp]
\centering
\includegraphics[width=\textwidth]{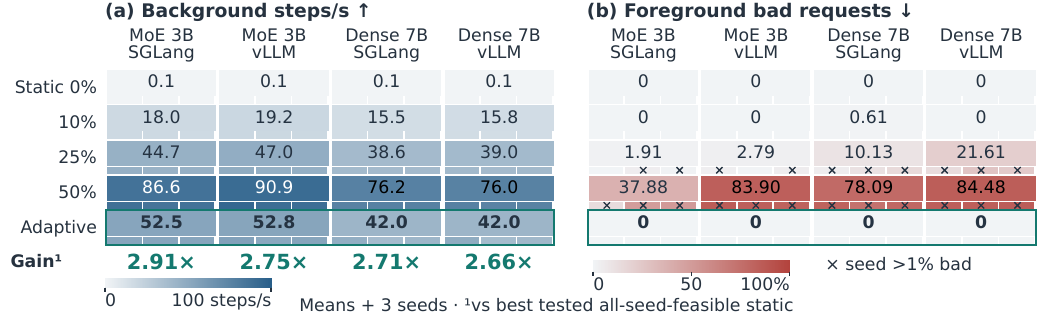}
\caption{Background progress and service risk across four systems. Numbers are three-seed means; strips show repeats, with crosses above 1\% bad requests. Gain is relative to the best measured static policy meeting the limit in every repeat.}
\Description{Background steps and bad-request rates for five resource policies across four systems, including all seeds.}
\label{fig:resource}
\end{figure*}
\begin{figure}[t]
\centering
\includegraphics[width=\columnwidth]{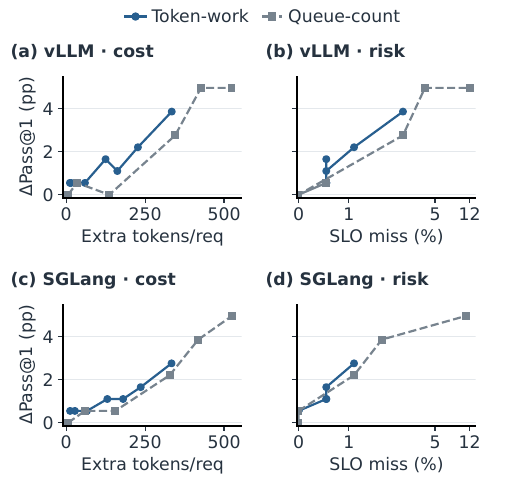}
\caption{Work-based versus count-based admission: all 26 settings, one 181-task run each. Left: cost--quality; right: risk--quality. A100 costs and latencies are measured; quality uses fixed-answer replay.}
\Description{Cost versus replay quality and SLO risk versus replay quality for two engines.}
\label{fig:work}
\end{figure}
\begin{figure}[t]
\centering
\includegraphics[width=\columnwidth]{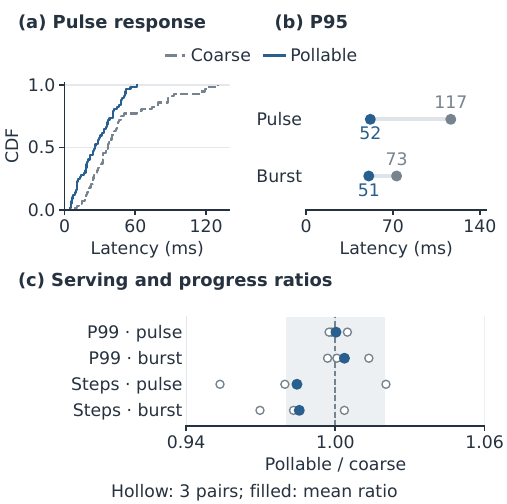}
\caption{Preemption under 5\%/0\% pulses and bursts, three runs per implementation, timed through execution acknowledgment. (a) Pulse CDF; (b) pulse/burst P95; (c) service P99 and background-rate ratios. Open/filled points show paired repeats/mean ratios; shading marks $\pm2\%$.}
\Description{Command actuation latency and system overhead for coarse and pollable waits.}
\label{fig:actuation}
\end{figure}
\paragraph{Why waiting helps}
The fresh case in Figure~\ref{fig:state_policy}(c) has 30 s of work, benefit rate 0.25, a back-loaded calendar, horizon 1020 s, and cold recovery. Costed Window advances 5 s at 200, checkpoints, and finishes with 25 s at 300; it publishes at 325.048 s for cost 170.092. Rollout starts at 300 and publishes at 330 for cost 99.988. Waiting saves 70.104 s-eq., loses 1.238 in benefits, and gains 68.866 net. With warm recovery, both publish at 30 s in the first window.

\subsection{Resource Allocation and Admission}
\label{sec:lending}
\paragraph{Adaptive shares}
Among static shares of 0, 10\%, 25\%, and 50\%, only shares up to 10\% satisfy the 1\% bad-request limit in every repeat across four systems. Adaptive allocation achieves 42.0--52.8 steps/s, or 2.66--2.91$\times$ the best measured feasible static policy, with no observed bad requests in 12 runs (Figure~\ref{fig:resource}). Expanding under low load and contracting under bursts uses capacity that static policies reserve for peaks. Appendix~\ref{app:tails} reports SGLang tails.

\begin{figure*}[!tbp]
\centering
\captionsetup{skip=6pt}
\includegraphics[width=\textwidth,trim={0 9.5bp 0 0},clip]{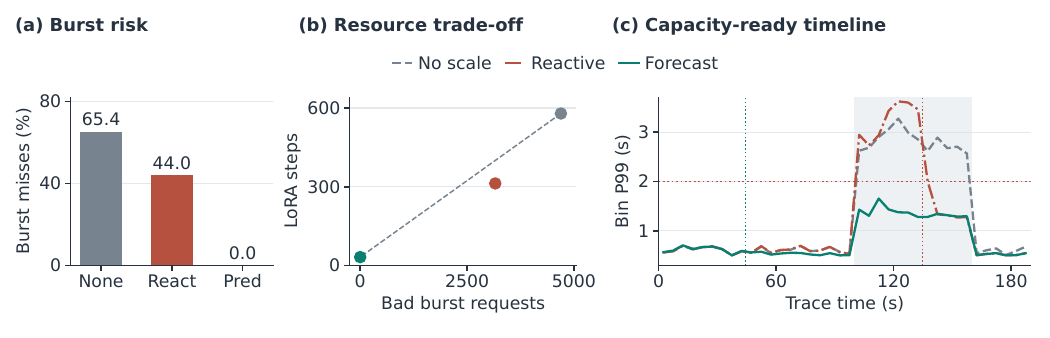}
\caption{vLLM capacity return on two A100s, three seeds per policy. None/React/Pred: no scaling, burst-triggered scaling, and a 90 s pre-alert. Panels show burst misses, progress--risk tradeoffs, and five-second P99/readiness for preselected seed 713, including client queuing.}
\Description{Service risk, evolution cost, and execution timelines for no scaling, reactive scaling, and advance-notice scaling.}
\label{fig:handoff}
\end{figure*}
\begin{figure}[t]
\centering
\captionsetup{skip=6pt}
\includegraphics[width=\columnwidth,trim={0 8bp 0 0},clip]{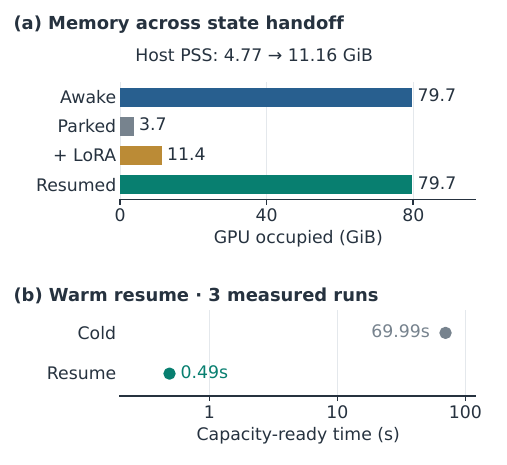}
\caption{Same-process SGLang recovery on one A100, three runs. GPU occupancy and cold/warm readiness (means and min--max); host PSS is separate. Every resumption completes a real request.}
\Description{Device memory in service, parked, parked-with-LoRA, and resumed states, with cold and warm recovery times.}
\label{fig:resident}
\end{figure}
\begin{figure*}[!tbp]
\centering
\includegraphics[width=\textwidth]{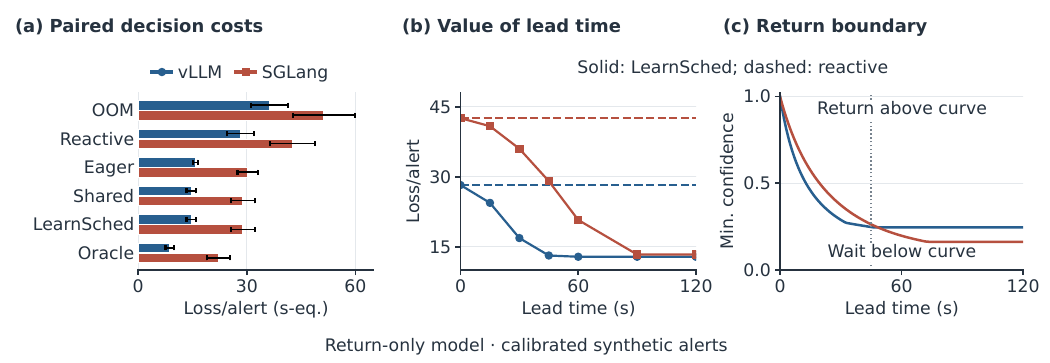}
\caption{Return decisions under synthetic alerts. (a) Mean losses of six rules and percentile ranges of seed means. (b) Seven lead times. (c) Return/wait boundaries. LearnSched estimates recovery cost per engine; Shared pools samples; Oracle knows events and actual durations. Loss is in s-eq./alert: one overflow-loss unit equals one second of equivalent evolution loss.}
\Description{Losses of six return policies, the value of lead time, and the LearnSched return/wait boundary.}
\label{fig:readiness_projection}
\end{figure*}
\paragraph{Work-based admission}
We sweep work budgets of 384--2048 and count thresholds of 0--12. Each setting dispatches 181 tasks as 40/101/40 requests at 2/6/2 requests/s. Quality comes from fixed-answer replay; costs and latencies come from real calls.
On vLLM, work budget 2048 and count threshold 4 both yield 2.21\% deadline misses. Their quality gains are 3.87 versus 2.76 percentage points at 334.1 versus 345.0 additional tokens/request. On SGLang, both yield 1.10\% misses, with gains of 2.76 versus 2.21 points at 333.6 versus 327.6 tokens/request.
Work-based admission gains more quality at the same risk, while cost advantages depend on the engine. Count threshold 12 raises miss rates to 12.15\% and 11.05\%. Figure~\ref{fig:work} shows both frontiers. A count limit treats requests with different prompt and output lengths alike; reserving estimated work makes the admission budget more sensitive to the load each request can impose. Engine-dependent costs reinforce the need to assess quality, resource use, and service risk together.

\subsection{Capacity Return and Recovery}
\label{sec:return}
\paragraph{Control response}
Pollable waits cut the P95 delay from command issuance to acknowledgment from 116.6 to 51.8 ms under pulses and from 73.0 to 50.6 ms under bursts (Figure~\ref{fig:actuation}). Every transition is acknowledged; neither implementation shows deadline misses. Burst service P99 is 0.562/0.564 s and background throughput 63.2/62.3 steps/s. Acknowledgment still precedes resource release and service readiness. In an independent SGLang sweep, 50 requests/s produces 86.8\% bad requests without failures, motivating deadline-risk triggers (Appendix~\ref{app:capacity}).

\paragraph{Early return and lost progress}
No scaling, reactive scaling, and advance-notice scaling yield mean burst miss rates of 65.4\%, 44.0\%, and zero, with P99 of 2.94, 3.31, and 1.38 s (Figure~\ref{fig:handoff}). Reactive capacity is ready 34.7 s after burst onset; advance-notice capacity is ready 56.3 s before it. Both start in 33--34 s, and all six handoff runs restore training exactly from final durable state.
Mean background progress falls from 579.0 to 312.7 and 32.0 steps. Under $E-wN_{\mathrm{bad}}$, reactive scaling beats no scaling only for $w\geq0.173$ and advance-notice scaling only for $w\leq0.089$, so it is never selected on this mean frontier. These runs establish the value and progress cost of advance notice when the burst is known. Uncertain alerts additionally require deciding whether to act on that notice.

\paragraph{Resident state}
Parking SGLang reduces GPU usage from 79.7 to 3.7 GiB; LoRA alongside it uses 11.4 GiB, and resumption restores 79.7 GiB (Figure~\ref{fig:resident}). All three runs resume the same process and complete a real request. Mean cold startup is 69.99 s, warm recovery 0.487 s, and recovery through first-request completion 0.520 s.
Device usage includes runtime reservations. Warm recovery holds 3.7 GiB on the GPU; host PSS rises from 4.77 to 11.16 GiB. These runs omit live prediction and value admission.

\begin{figure*}[!tbp]
\centering
\includegraphics[width=\textwidth]{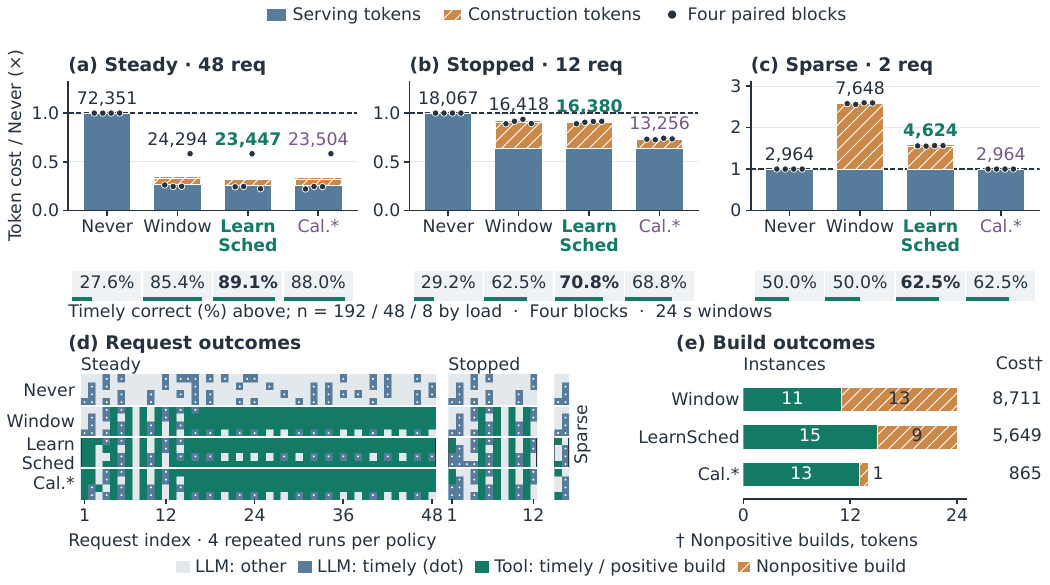}
\caption{Demand persistence and investment returns. (a--c) Total tokens, cost breakdowns, and timely correctness. Points show cost ratios from four historical blocks; quality uses revised summaries with denominators of 192, 48, and 8. (d) The 992 request slots arranged by aggregate counts. (e) Revised construction classes and costs of nonpositive-return builds. Cal.* knows the future calendar.}
\Description{Historical costs, revised quality summaries, illustrative request classifications, and construction outcomes under three reuse windows.}
\label{fig:causal}
\end{figure*}
\begin{figure*}[!tbp]
\centering
\includegraphics[width=\textwidth]{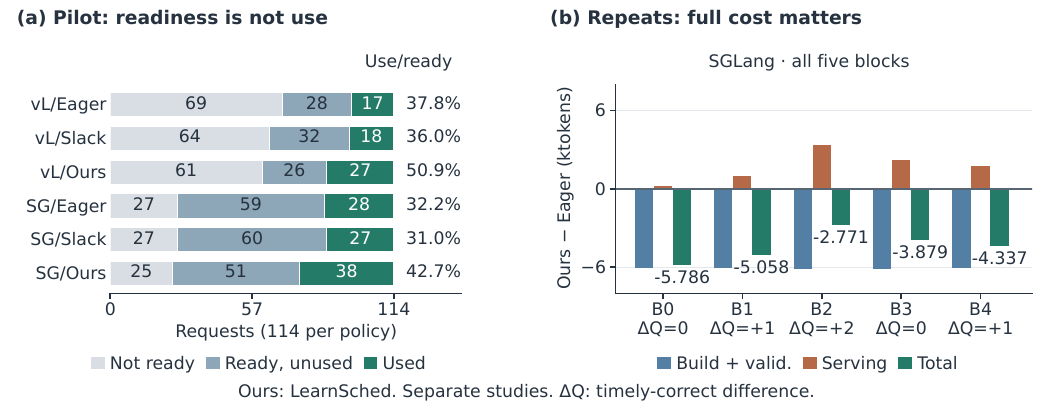}
\caption{Effective guidance use and conditional net cost; Ours denotes LearnSched. (a) Three groups per engine, each with 114 requests; ratios are use/ready. (b) Cost differences from Eager in five SGLang blocks: construction differences retain the original records; service differences are scaled by $\alpha=0.561$.}
\Description{Guidance visibility and use counts, with the net cost breakdown of LearnSched relative to Eager under the stated service-cost coefficient.}
\label{fig:mechanism}
\end{figure*}
\begin{figure}[t]
\centering
\includegraphics[width=\columnwidth]{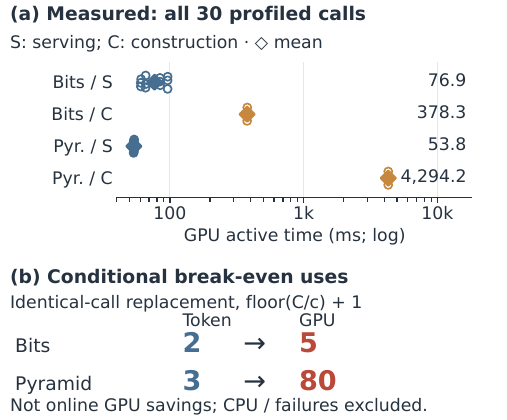}
\caption{Device pricing and break-even reuse. Top: all 30 profiled calls, with diamonds marking means. Bottom: minimum uses needed to offset construction costs, assuming no additional cost per use.}
\Description{Service and construction GPU active times for Bits and Pyramid, and their conditional break-even reuse counts under token and GPU pricing.}
\label{fig:calibration}
\end{figure}
\paragraph{Acting on uncertain alerts}
Figure~\ref{fig:readiness_projection} combines measured recovery samples with calibrated synthetic alerts to isolate this decision. Early return trades the service loss avoided under a true burst against the evolution cost paid under a false alarm. The resulting threshold depends on both alert confidence and lead time; Appendix~\ref{app:risk} gives the derivation, distributions, and assumed prices.
On vLLM, reactive return, unconditional early return, and LearnSched incur 28.16, 15.76, and 14.61 s-eq./alert; SGLang yields 42.48, 30.21, and 28.87. Lead time explains most of the gain, with selective return adding savings. Pooling recovery samples across engines gives 14.63/28.90, close to engine-specific estimates. Thus, predicting earlier and pricing false alarms matter more here than refining recovery estimates. This return-only model uses synthetic events; it neither establishes online prediction gains nor jointly prices residency.

\subsection{Capability Investment and Payback}
\label{sec:investment}
\paragraph{Demand persistence}
Never does not invest; Window checks the remaining window; LearnSched's value layer also estimates amortization. All share construction and public qualification. Under sustained demand, LearnSched spends 23447 tokens for 171/192 timely correct answers, versus Never's 72351 for 53/192 and Window's 24294 for 164/192 (Figure~\ref{fig:causal}). Historical admission decisions match those of Window; 745 of the 847-token difference comes from service costs, so this difference does not establish improved investment selection.

Under stopped demand, LearnSched and Window spend 16380/16418 tokens for 34/48 versus 30/48 timely correct answers. Under sparse demand, they spend 4624/7648 for 5/8 versus 4/8. LearnSched's extra timely answer over Never costs 1660 tokens, requiring $W>1660$ for positive incremental utility. The construction outcomes in Figure~\ref{fig:causal}(e) expose another cost: LearnSched and Window have 9 and 13 nonpositive-return builds, costing 5649 and 8711 tokens. Qualification alone does not ensure payback. Cal.* achieves the same sparse-demand quality as LearnSched with 2964 tokens, illustrating the remaining value of future demand information. Further ledgers appear in Appendix~\ref{app:investment}.

\paragraph{From publication to effective reuse}
Guidance makes the distinction between availability and use explicit (Figure~\ref{fig:mechanism}(a)). LearnSched has 53 ready opportunities and 27 uses on vLLM, and 89 ready opportunities and 38 uses on SGLang. Its use/ready ratios are 50.9\% and 42.7\%, compared with Eager's 37.8\% and 32.2\%. Many requests still do not use a visible artifact. Counting every post-publication request as reuse overestimates amortization; use probability and per-use quality need separate estimates.

Construction savings must also survive the use path. In the conditional five-block SGLang ledger, service overhead offsets 28.24\% of the 30420-token construction/validation savings, leaving 21830.529 tokens net (Figure~\ref{fig:mechanism}(b)). Every block saves cost, with revised timely-correct differences of 0, 1, 2, 0, and 1. These results use the stated service-cost coefficient $\alpha=0.561$ and separately measured use counts. They support accounting for both construction and service, rather than inferring net value from fewer builds alone.

\paragraph{Resource pricing}
Thirty profiler-on/off pairs across 10 prompts have equal paired output usage. The 30 profiled calls comprise 20 service and 10 construction calls. Mean service/build GPU activity is 76.9/378.3 ms for Bits and 53.8/4294.2 ms for Pyramid. Assuming no additional use cost, positive payback requires $\lfloor C/c\rfloor+1$ uses: 2 and 3 under token pricing, but 5 and 80 under GPU pricing (Figure~\ref{fig:calibration}). GPU activity is the union of kernel, copy, and memset intervals; CPU costs are separate.
In an independent sustained batch, TokenROI uses 73.76\% fewer tokens than Never but 86.23\% more GPU activity. GPUROI saves 71.25\% of that activity but loses 32 timely correct answers. Under stopped demand, it saves 86.16\% with unchanged timely quality. Resource savings and quality effects require joint assessment.

\section{Discussion and Related Work}
\subsection{State Management and Scheduling Complexity}
State management changes both feasible windows and plan values. With expensive cold recovery, checkpointing and segmented execution can incur substantial handoff costs, making a later, longer window more valuable. Warm recovery can bring a window policy close to the value of more complex planning.
Runtime design and scheduling therefore interact: retaining state must be valued through faster return, less repeated work, and residency costs. Planning and prediction should focus where they change actions.

\subsection{Capability Maintenance and Long-Term Value}

The current evaluation focuses on capability construction and subsequent reuse, without systematically accounting for maintenance costs such as version updates, failure detection, and revalidation. As task requirements or tool interfaces change, published capabilities may become less effective, so short-term cost savings may not translate into long-term net value. Future work should assess how ongoing maintenance and declining benefits affect investment decisions over the capability lifecycle.

\subsection{Related Work}
Jalape\~no selects recompilation based on expected future execution benefits~\cite{jikesrvm}; computation selection and metareasoning study when to invest further computation~\cite{metaselection,bmps}.
We include qualification, actual use, and service-recovery paths in remaining investment value, and use classical rollout~\cite{rollout1997} to compare capacity-constrained execution plans.

Ekya coordinates inference and retraining, RECL reuses historical models, and LeMix jointly schedules LLM training and inference~\cite{ekya,recl,lemix}.
MetaClaw learns during inactivity and calendar windows~\cite{metaclaw}. Yan et al. propose a cost-aware evolution control plane with multiple actions and instantiate online weight updates through AReaL2.0~\cite{evolutionplane}.
\system focuses on break-even value, durable progress, and capacity-return timing for discrete capabilities, evaluating plan differences under the same information and execution method.

Gandiva time-slices and migrates training jobs, Tiresias targets job completion time, and AntMan dynamically shares spare GPU resources~\cite{gandiva,tiresias,antman}.
Salus provides fast job switching and memory sharing, while REEF supports microsecond-scale preemption for DNN inference~\cite{salus,reef}; these mechanisms inform the cost of lending and returning capacity.
Clockwork uses predictable execution to meet inference latency targets~\cite{clockwork}. Orca schedules at iteration granularity, Sarathi-Serve uses chunked prefills, and DistServe separates prefill and decoding to improve LLM serving under latency constraints~\cite{orca,sarathi,distserve}.

PagedAttention and SGLang improve execution efficiency~\cite{vllm,sglang}, while Reflexion and Self-Refine provide feedback-based improvement mechanisms~\cite{reflexion,selfrefine}. Voyager accumulates reusable executable skills~\cite{voyager}.
Recent advances extend these mechanisms to evolutionary program optimization and self-directed model adaptation~\cite{alphaevolve,seal}, as well as skill--tool co-evolution and adaptive workflow evolution~\cite{skillsmith,evomas}. DGM and Hyperagents investigate agent self-modification, with the latter also making the meta-level improvement procedure itself editable~\cite{dgm,hyperagents}.
These advances expand both the capabilities available for reuse and the mechanisms that construct them. Investment scheduling must therefore re-estimate construction and validation costs, expected reuse benefits, and capacity-return overhead as the system evolves.

\FloatBarrier
\section{Conclusion}
\system treats online evolution as a compute investment with delayed returns and state-dependent switching costs. Counterfactual rollout compares progress and waiting under capacity-return constraints. Measurements and finite-model results show that waiting can avoid costly cold handoffs, while warm recovery reduces this advantage. Effective evolution scheduling must jointly account for future benefits and timely capacity return.

\FloatBarrier
\clearpage
\newpage
\bibliographystyle{ACM-Reference-Format}
\bibliography{references}

\clearpage
\appendix
\setcounter{figure}{0}
\renewcommand{\thefigure}{\thesection.\arabic{figure}}
\renewcommand{\theHfigure}{appendix.\thesection.\arabic{figure}}
\setcounter{equation}{0}
\renewcommand{\theequation}{\thesection.\arabic{equation}}
\renewcommand{\theHequation}{appendix.\thesection.\arabic{equation}}
\setcounter{table}{0}
\renewcommand{\thetable}{\thesection.\arabic{table}}
\renewcommand{\theHtable}{appendix.\thesection.\arabic{table}}

\section{Service Risk: Measurement and Interpretation}
\label{app:service_details}
The service experiments quantify the capacity cost of background work. This appendix specifies how latency, failures, and deadline violations enter that measurement, and distinguishes the burst experiments from the independent capacity sweep. These observations support the capacity-return constraint in Section~\ref{sec:model}; they do not establish a deadline guarantee under arbitrary arrivals.

\subsection{Response Tails and Failure Accounting}
\label{app:tails}
\begin{figure}[!htbp]
\centering
\includegraphics[width=\columnwidth]{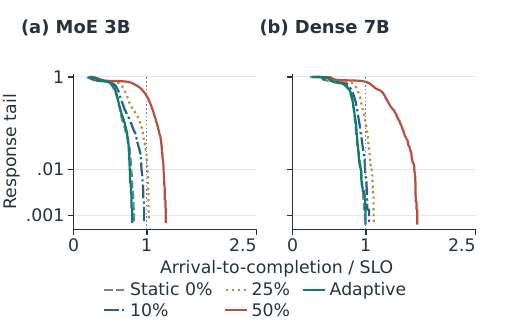}
\caption{Successful-response latency distributions for the two SGLang models and five resource policies. Latency includes waiting from scheduled arrival and is divided by the model's SLO. Failed requests are excluded from these curves and retained in the bad-request denominator.}
\Description{Normalized successful-response latency distributions under adaptive and four static resource policies for two SGLang models. The curves condition on successful completion.}
\label{fig:tails_sglang}
\end{figure}

Figure~\ref{fig:tails_sglang} supplements Section~\ref{sec:lending}. The SGLang experiments cross two models and five resource policies, with three repetitions per setting: 30 runs contain 14310 requests, of which 14308 succeed. The corresponding SLOs are 1.2 s for Granite-3B-A0.8B and 1.8 s for Qwen2.5-Coder-7B. Arrivals follow the low--high--low workload in Section~\ref{sec:setup}.

\paragraph{Latency origin and normalization}
For a successful request $i$, let $a_i$ be its scheduled arrival, $f_i$ its completion time, and $D_i=d_i-a_i>0$ its latency budget. The normalized latency is $x_i=(f_i-a_i)/D_i$. Measuring from scheduled arrival includes client-side waiting before dispatch, so delaying dispatch cannot hide a deadline violation. Normalization puts the SLOs on a common axis: $x_i>1$ denotes lateness, while absolute latency still depends on the model.

\paragraph{Conditional tails versus total service risk}
Partition the $N>0$ requests into disjoint failure and success sets $\mathcal F$ and $\mathcal S$, with $N_f=|\mathcal F|$ and $N_s=|\mathcal S|=N-N_f$. For $N_s>0$, define the empirical conditional tail
\begin{equation}
 \begin{aligned}
 \widehat S_s(x)
 &=\frac{|\{i\in\mathcal S:x_i>x\}|}{|\mathcal S|}\\
 &=\frac{1}{N_s}\sum_{i\in\mathcal S}\mathbf 1\{x_i>x\}.
 \end{aligned}
 \label{eq:success_tail}
\end{equation}
Let $N_\ell=|\{i\in\mathcal S:x_i>1\}|$ count late successes, and let $\widehat f=N_f/N$. A request is bad if it fails or succeeds late. Because these events are disjoint, their counts give the exact decomposition
\begin{equation}
 \begin{aligned}
 \widehat b
 &=\frac{|\mathcal F|+|\{i\in\mathcal S:x_i>1\}|}{N}\\
 &=\frac{N_f+N_\ell}{N}
  =\frac{N_f}{N}+\frac{N_s}{N}\frac{N_\ell}{N_s}\\
 &=\widehat f+(1-\widehat f)\widehat S_s(1).
 \end{aligned}
 \label{eq:bad_decomposition}
\end{equation}
If $N_s=0$, the conditional tail is undefined and the count formula directly gives $\widehat b=1$. A successful-response curve alone cannot recover total risk without the failure fraction. Both failures in the SGLang experiments therefore remain in the denominator of Figure~\ref{fig:resource}, although they have no finite successful-completion latency in Figure~\ref{fig:tails_sglang}.

The identity applies within a run or to consistently pooled counts. Averaging per-run rates gives each repetition equal weight; pooling requests weights runs by their request counts. These summaries should not be interchanged when denominators differ. The three repetitions reveal observed variability, but an absence of bad requests in this finite workload does not establish zero risk for unseen bursts.

\subsection{Independent Capacity Sweep}
\label{app:capacity}
\begin{figure}[!htbp]
\centering
\includegraphics[width=\columnwidth]{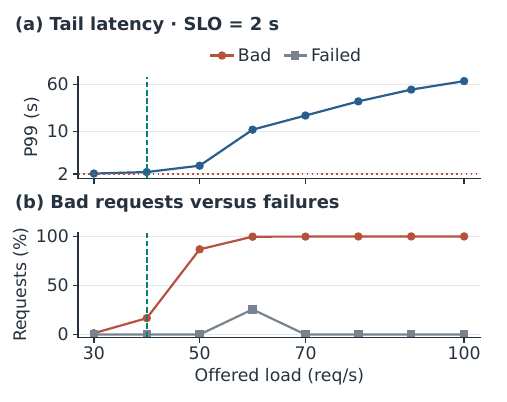}
\caption{Independent SGLang capacity sweep with open-loop Poisson arrivals and a two-second SLO. Each rate has one run. The vertical line marks 40 requests/s; the rise in deadline violations precedes observed request failures.}
\Description{P99 latency, bad-request rate, and failure rate under increasing offered load. Deadline violations rise before failures appear.}
\label{fig:capacity}
\end{figure}

Figure~\ref{fig:capacity} sweeps offered load for SGLang with Granite 3B on one A100. It uses open-loop Poisson arrivals, a two-second SLO, and one run per rate. The open-loop workload does not reduce arrivals when responses slow down, allowing queueing pressure to accumulate. Its SLO and arrival process differ from the preceding burst experiment, so the curves are not additional repetitions of that experiment.

Increasing arrivals from 40 to 50 requests/s raises the bad-request rate from 16.7\% to 86.8\% and P99 from 2.14 to 2.73 s, without failures at either rate. Failures first appear at 60 requests/s. Deadline violations thus expose a loss of useful capacity before the runtime begins failing requests. Failure-only control misses this earlier degradation.

This sweep identifies an observed operating boundary for the specified workload, rather than a workload-independent capacity constant. One run per rate cannot characterize tail uncertainty or establish the exact saturation point. The scheduling implication is narrower: capacity admission and return should account for deadline pressure, including waiting outside the serving engine.

\FloatBarrier
\section{Investment Ledgers and Price Sensitivity}
\label{app:investment}
\setcounter{figure}{0}
\setcounter{equation}{0}
\setcounter{table}{0}
The ledgers separate the cost of constructing a capability from the cost and quality of using it. They complement the plan comparison in Section~\ref{sec:state_eval}: a lower realized bill can arise from a different service path even when admission decisions are unchanged. The historical costs and revised quality summaries support aggregate comparisons.

\subsection{Demand Persistence and Incremental Value}
Table~\ref{tab:investment_ledger} collects the comparisons underlying Figure~\ref{fig:causal}. All token totals include charged construction and validation, including unsuccessful or unused work, as specified by Equation~\eqref{eq:ledger}. Timely correctness counts answers that meet both the correctness and deadline requirements.

\begin{table}[!htbp]
\centering
\caption{Reported tool ledgers by demand pattern. Costs are historical token totals; timely-correct counts use the revised summaries. Cal.* knows future arrivals.}
\label{tab:investment_ledger}
\small
\setlength{\tabcolsep}{4pt}
\begin{tabular}{@{}llrr@{}}
\toprule
Demand & Policy & Tokens & Timely correct \\
\midrule
Sustained & Never & 72351 & 53/192 \\
 & Window & 24294 & 164/192 \\
 & LearnSched & 23447 & 171/192 \\
\midrule
Stopped & Window & 16418 & 30/48 \\
 & LearnSched & 16380 & 34/48 \\
 & Cal.* & 13256 & 33/48 \\
\midrule
Sparse & Window & 7648 & 4/8 \\
 & LearnSched & 4624 & 5/8 \\
 & Cal.* & 2964 & 5/8 \\
\bottomrule
\end{tabular}
\end{table}

Under sustained demand, LearnSched saves 67.59\% of tokens relative to Never and 3.49\% relative to Window. Its seven additional timely correct answers over Window correspond to 3.65 percentage points. However, LearnSched and Window make identical historical admission decisions, and 745 of their 847-token cost difference comes from service. These totals demonstrate a realized cost--quality difference; they do not establish better plan selection.

For a policy $P$, let $Q_P$ count timely correct answers, $C_P$ be the charged token cost, and $W\geq0$ value one timely correct answer in token-cost units. On a common workload, $U_P(W)=WQ_P-C_P$ gives
\begin{equation}
 \begin{aligned}
 \Delta U_{P,Q}(W)
 &=U_P(W)-U_Q(W)\\
 &=(WQ_P-C_P)-(WQ_Q-C_Q)\\
 &=W(Q_P-Q_Q)-(C_P-C_Q).
 \end{aligned}
 \label{eq:ledger_delta}
\end{equation}
For LearnSched relative to Cal.*, substitution of the stopped and sparse ledgers yields
\begin{equation}
 \begin{aligned}
 \Delta U_{\rm stopped}(W)
 &=W(34-33)-(16380-13256)\\
 &=W-3124,\\
 \Delta U_{\rm sparse}(W)
 &=W(5-5)-(4624-2964)\\
 &=-1660.
 \end{aligned}
 \label{eq:ledger_thresholds}
\end{equation}
Under stopped demand, LearnSched's incremental utility is therefore positive if and only if $W>3124$; equality gives a tie. Under sparse demand, equal aggregate quality and strictly greater cost favor Cal.* for every common $W$. Cal.* has privileged knowledge of future arrivals, so neither comparison holds information access fixed.

Qualification also differs from amortization. Across the revised workload classifications, LearnSched and Window incur 5649 and 8711 tokens on nonpositive-return builds, respectively, a 35.15\% difference. These costs remain in the ledger even when a candidate passes public tests. Passing those tests establishes eligibility for publication; positive return requires sufficient reuse at a worthwhile unit value.

\subsection{Conditional Guidance Costs}
Figure~\ref{fig:mechanism}(b) preserves construction and validation differences from the five historical SGLang blocks, while multiplying service differences by $\alpha_0=0.561$. Let $S_b=30420$ be the aggregate construction saving and $D_s$ the unscaled additional service cost. Holding these quantities fixed, the saving at a service-price coefficient $\alpha\geq0$ follows from
\begin{equation}
 \begin{aligned}
 D_s&=\frac{8589.471}{\alpha_0}
     =\frac{8589.471}{0.561}=15311,\\
 S(\alpha)&=S_b-\alpha D_s\\
           &=30420-15311\alpha.
 \end{aligned}
 \label{eq:guidance_sensitivity}
\end{equation}
At the reported coefficient, $S(0.561)=21830.529$ token-cost units: service offsets 28.24\% of the construction saving. The revised timely-correct differences are $0,1,2,0,1$, giving four additional answers in aggregate. Consequently,
\begin{equation}
 \begin{aligned}
 \Delta U(W,\alpha)
 &=W\sum_{j=1}^{5}\Delta Q_j+S(\alpha)\\
 &=W(0+1+2+0+1)+S(\alpha)\\
 &=4W+30420-15311\alpha.
 \end{aligned}
 \label{eq:guidance_utility}
\end{equation}
Since $D_s=15311>0$, the sign of incremental utility has the equivalent conditions
\begin{equation}
 \begin{aligned}
 \Delta U(W,\alpha)>0
 &\iff 15311\alpha<30420+4W\\
 &\iff \alpha<\frac{30420+4W}{15311}.
 \end{aligned}
 \label{eq:guidance_boundary}
\end{equation}
At $W=0$, the boundary is $30420/15311\approx1.9868$; equality gives a tie. These sensitivities hold ledger outcomes fixed. They do not measure new operating points or account for behavioral, quality, or reuse changes induced by new prices.

The use counts in Figure~\ref{fig:mechanism}(a) come from a separate experiment. They demonstrate the distinction between artifact availability and actual use, but cannot be joined request by request to the conditional cost ledger. Per-use savings remain unidentified, and missing usage remains unknown.

\FloatBarrier
\section{Return, Stopping, and Benefit Timing}
\label{app:risk}
\setcounter{figure}{0}
\setcounter{equation}{0}
\setcounter{table}{0}
This appendix examines three decision boundaries: returning capacity after an alert, stopping an in-flight computation, and recognizing benefits after publication. Each is an instance of valuing remaining work. The experiments use different workloads and information, so their results characterize these boundaries separately rather than a single integrated online policy.

\subsection{Returning Capacity Under an Alert}
Restrict Equation~\eqref{eq:bellman} to two feasible actions: return early or wait. Fix the information $\mathcal I$ available at the decision, including warning lead time $L$, and let $E$ denote a true burst, with $p=\Pr(E\mid\mathcal I)\in[0,1]$. All expectations below condition on this same information; write $\mathbb E_{\mathcal I}[\cdot]=\mathbb E[\cdot\mid\mathcal I]$.

Let $\ell_w$ and $\ell_a$ be integrable remaining losses under waiting and early return. Subtract a common event-dependent baseline from both losses so that waiting on a false alarm has conditional mean zero; this preserves their expected difference. Under a true burst, the conditional means are $C_w$ and $C_a(L)$. Waiting includes the better feasible continuation of reactive scaling and no scaling under the information then available. Under a false alarm, early return has incremental mean cost $C_f>0$. The law of total expectation gives
\begin{equation}
 \begin{aligned}
 Q_w&=\mathbb E_{\mathcal I}[\ell_w\mathbf 1_E]
       +\mathbb E_{\mathcal I}[\ell_w\mathbf 1_{E^c}]\\
    &=pC_w+(1-p)\,0=pC_w,\\
 Q_a&=\mathbb E_{\mathcal I}[\ell_a\mathbf 1_E]
       +\mathbb E_{\mathcal I}[\ell_a\mathbf 1_{E^c}]\\
    &=pC_a(L)+(1-p)C_f.
 \end{aligned}
 \label{eq:return_scores}
\end{equation}
Here $Q_w$ and $Q_a$ are losses to minimize. Conditional costs include every remaining consequence of the specified continuations; their dependence on $\mathcal I$ is suppressed. At $p=0$ or $p=1$, the zero-probability branch contributes zero. Previously paid costs are common to both actions and cancel.

Write $B(L)=C_w-C_a(L)$ for the conditional benefit of acting early. For $B(L)>0$, early return is strictly preferred if and only if
\begin{equation}
 \begin{aligned}
 &Q_a<Q_w\\
 &\iff pC_a(L)+(1-p)C_f<pC_w\\
 &\iff pB(L)>(1-p)C_f\\
 &\iff p\bigl[B(L)+C_f\bigr]>C_f\\
 &\iff p>p^*(L):=\frac{C_f}{B(L)+C_f}.
 \end{aligned}
 \label{eq:return_gate}
\end{equation}
Division preserves the inequality because $B(L)+C_f>0$, and $p^*(L)\in(0,1)$. If $B(L)\leq0$, then $pB(L)-(1-p)C_f\leq0$ for every $p\in[0,1]$, so early return cannot strictly improve the loss. Ties retain waiting.

\paragraph{Sensitivity and calibration}
For $B>0$ and $C_f>0$, apply the quotient rule while holding the other parameter fixed. The threshold decreases with avoidable loss and increases with false-alarm cost:
\begin{equation}
 \begin{aligned}
 \frac{\partial p^*}{\partial B}
 &=\frac{0\cdot(B+C_f)-C_f}{(B+C_f)^2}\\
 &=-\frac{C_f}{(B+C_f)^2}<0,\\
 \frac{\partial p^*}{\partial C_f}
 &=\frac{(B+C_f)-C_f}{(B+C_f)^2}\\
 &=\frac{B}{(B+C_f)^2}>0.
 \end{aligned}
 \label{eq:return_sensitivity}
\end{equation}
Longer lead time lowers the threshold only insofar as it increases $B(L)$. Faster recovery can instead reduce the extra loss avoided by advance notice, making early intervention less valuable. Recovery latency should thus change the action value, not merely a fixed scheduling margin.

Keep $B>0$ and $C_f>0$ fixed, and suppose $\widehat p\in[0,1]$ satisfies $|p-\widehat p|\leq\epsilon_p$. Define the true advantage of early return as $D(p)=Q_w-Q_a=pB-(1-p)C_f$. Expanding the difference and applying the probability-error bound gives
\begin{equation}
 \begin{aligned}
 D(p)-D(\widehat p)
 &=(p-\widehat p)B+(p-\widehat p)C_f\\
 &=(p-\widehat p)(B+C_f),\\
 |D(p)-D(\widehat p)|
 &=|p-\widehat p|(B+C_f)\\
 &\leq\epsilon_p(B+C_f),\\
 D(p)&\geq D(\widehat p)-\epsilon_p(B+C_f).
 \end{aligned}
 \label{eq:return_calibration}
\end{equation}
Thus $D(\widehat p)=\widehat pB-(1-\widehat p)C_f>\epsilon_p(B+C_f)$ is sufficient for $D(p)>0$ and lower true loss. This conservative condition excludes cost-model errors. Synthetic events follow their stated probabilities, which does not validate conditional error bounds for a live predictor.

\subsection{Synthetic Alert Protocol and Its Scope}
Figure~\ref{fig:readiness_projection} combines measured readiness and failed-wait samples with synthetic alerts. Alert probabilities follow $p\sim\mathrm{Beta}(2,2)$, and event occurrence is sampled with probability $p$. Duration, lead time, and overflow rate follow lognormal distributions with the parameters in Table~\ref{tab:alert_parameters}. To preserve an arithmetic mean $m>0$, write $X=e^{\mu+\sigma Z}$ with $Z\sim\mathcal N(0,1)$ and $\sigma\geq0$. Completing the square in the normal-density integral gives
\begin{equation}
 \begin{aligned}
 \mathbb E[X]
 &=\frac{e^\mu}{\sqrt{2\pi}}
   \int_{\mathbb R}e^{\sigma u-u^2/2}\,du\\
 &=e^{\mu+\sigma^2/2}
   \int_{\mathbb R}\frac{e^{-(u-\sigma)^2/2}}{\sqrt{2\pi}}\,du\\
 &=e^{\mu+\sigma^2/2}=m,\\
 \mu&=\log m-\sigma^2/2.
 \end{aligned}
 \label{eq:lognormal_mean}
\end{equation}
The shifted normal density integrates to one; taking logarithms is valid because $m>0$.

\begin{table}[!htbp]
\centering
\caption{Synthetic alert parameters. Means are in the original scale; standard deviations are in log space. These distributions are modeling inputs.}
\label{tab:alert_parameters}
\small
\begin{tabular}{@{}lrr@{}}
\toprule
Quantity & Arithmetic mean & Log-space s.d. \\
\midrule
Burst duration & 90 s & 0.65 \\
Warning lead time & 45 s & 0.35 \\
Overflow rate & 1.2 & 0.50 \\
\bottomrule
\end{tabular}
\end{table}

Handoff loss is 13.84 s-eq., and competing policies share the same sampled events. One overflow-loss unit is priced as one second of equivalent evolution loss. Pairing holds the realized event sequence fixed when comparing decisions; it does not remove uncertainty in the underlying recovery measurements or validate the assumed loss conversion.

Oracle knows event occurrence and actual duration, achieving 8.74/22.16 s-eq./alert on vLLM/SGLang. It is an information benchmark rather than an implementable predictor. LearnSched's corresponding losses are 14.61/28.87, versus 15.76/30.21 for unconditional early return. These results isolate the value of selective action after notice is available. Online prediction accuracy and warm-state memory costs remain unevaluated.

\subsection{Budget Stopping During Execution}
\begin{figure}[!htbp]
\centering
\includegraphics[width=\columnwidth]{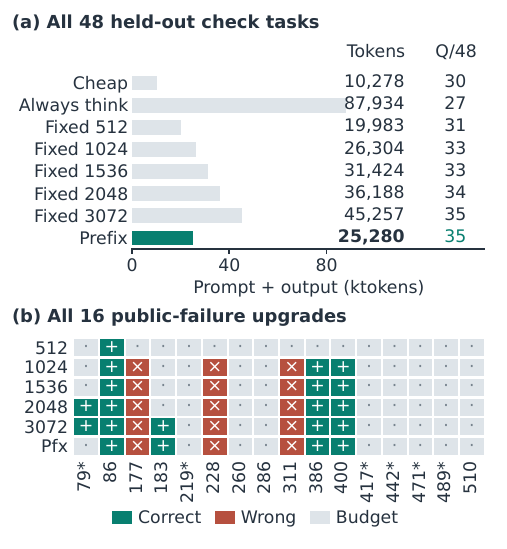}
\caption{Cost and quality under stage stopping. (a) Aggregate token costs and correct counts for eight policies; Prefix's 35/48 uses the revised summary. (b) Recorded outcomes for 16 upgrades after failed public tests, including six format-triggered stops. The per-task matrix remains only partly reconciled with the revised totals.}
\Description{Aggregate stopping-policy results and partially revised task records, with correct, incorrect, budget-exhausted, and format-stopped outcomes distinguished.}
\label{fig:prefix}
\end{figure}

Figure~\ref{fig:prefix} replays a Prefix rule frozen on development data against existing MBPP records. Failure of a public test on the cheap path triggers an upgrade. At step 512, the rule inspects the first 64 reasoning characters: a prefix of \texttt{Thinking Process:} stops execution; otherwise, execution continues to step 1536. This format is a model-specific proxy, rather than a direct estimate of the value of further computation.

In the revised summary, Prefix spends 25280 tokens for 35/48 correct answers, matching the fixed 3072-step budget's aggregate correctness at 44.14\% lower cost than its 45257 tokens. The fixed 1024-step budget spends 26304 tokens for 33/48; Cheap and the fixed 512-step budget answer 30 and 31 tasks correctly. These counts measure correctness; Appendix~\ref{app:investment} instead uses timely correctness.

At an intermediate history $h$, compare immediate stopping with a specified feasible continuation policy. Let $Y_c,Y_s\in\{0,1\}$ denote their eventual correctness and $C_c,C_s$ their remaining charged costs, including any required capacity return. Assume finite conditional expected costs, and let $W\geq0$ value a correct answer. Both options have already paid $C_{\rm paid}(h)$. With $\Delta Q_h=Y_c-Y_s$ and $\Delta C_h=C_c-C_s$, linearity of conditional expectation gives
\begin{equation}
 \begin{aligned}
 U_c(h)&=\mathbb E[WY_c-C_c\mid h]-C_{\rm paid}(h),\\
 U_s(h)&=\mathbb E[WY_s-C_s\mid h]-C_{\rm paid}(h),\\
 U_c(h)-U_s(h)
 &=\mathbb E[WY_c-C_c\mid h]
   -\mathbb E[WY_s-C_s\mid h]\\
 &=W\mathbb E[Y_c-Y_s\mid h]
   -\mathbb E[C_c-C_s\mid h]\\
 &=W\mathbb E[\Delta Q_h\mid h]
   -\mathbb E[\Delta C_h\mid h].
 \end{aligned}
 \label{eq:stopping_value}
\end{equation}
Consequently, this continuation is strictly preferable exactly when $W\mathbb E[\Delta Q_h\mid h]>\mathbb E[\Delta C_h\mid h]$. Previously consumed tokens cancel, while future costs remain. Global optimality would additionally require optimizing the continuation at every future history. Prefix uses a stopping signal without estimating these conditional values.

Equal aggregate correctness also does not imply equal correctness on every task: Prefix truncates task 79, which the longer budget answers correctly. In a historical transfer across five task families, its cost for the same three skills is 5.26\% higher than the fixed 1024-step budget. The revised 35/48 total has not been fully reconciled with per-task records. The evidence therefore supports a workload-specific aggregate tradeoff, not universal dominance or a validated stopping rule across models.

\subsection{Benefit Accumulation After Publication}
\begin{figure*}[!tbp]
\centering
\includegraphics[width=\textwidth]{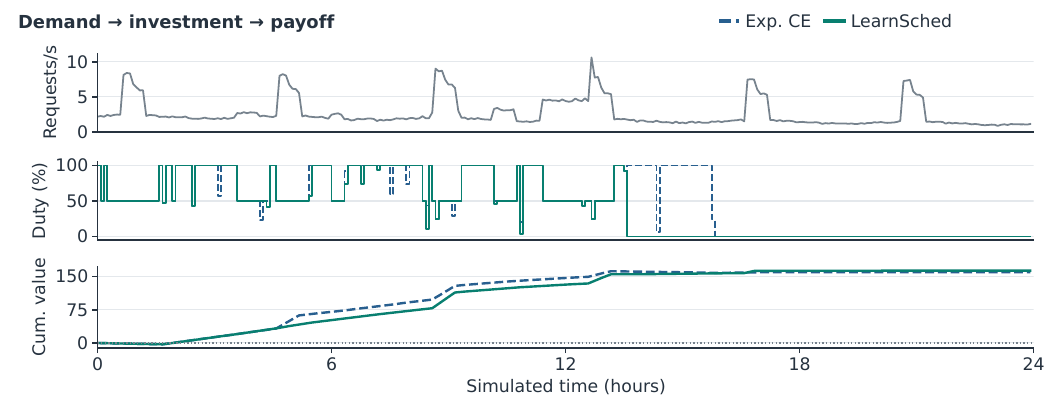}
\caption{Compute expenditure and delayed benefit under synthetic demand. The displayed run uses seed 261000, 24 hours, 24 candidates, and a 12 GPU-hour budget. Exp. CE and the conservative rule share the simulator. Demand, resource shares, and cumulative values are simulated; the trace is not an online deployment measurement.}
\Description{Simulated demand, background resource shares, and cumulative net value under two policies with common workload and budget parameters.}
\label{fig:timeline}
\end{figure*}

Figure~\ref{fig:timeline} uses a separate simulator with stable, single-burst, recurrent, and decaying demand in five-minute slots. The displayed run fixes seed 261000, a 24-hour horizon, 24 candidates, and a 12 GPU-hour budget. Benefits accrue after the completion slot and decay with artifact age. Costs, completion delays, and queue replacement are shared.

For the finite candidate set $\mathcal J$, let $k_j$ be candidate $j$'s completion slot, $b_{j,k}$ its simulated benefit in slot $k$, and $c_k$ the cost charged in that slot. Set $k_j=\infty$ for unfinished candidates. With initial value $G(-1)=0$, summing slot increments and exchanging finite sums yields
\begin{equation}
 \begin{aligned}
 G(K)
 &=\sum_{k=0}^{K}\bigl[G(k)-G(k-1)\bigr]\\
 &=\sum_{k=0}^{K}
   \left[\sum_{j\in\mathcal J}\mathbf 1\{k>k_j\}b_{j,k}-c_k\right]\\
 &=\sum_{\substack{j\in\mathcal J:\,k_j<K}}
   \sum_{k=k_j+1}^{K}b_{j,k}-\sum_{k=0}^{K}c_k.
 \end{aligned}
 \label{eq:timeline_accounting}
\end{equation}
Completion slots are nonnegative integers when finite, and an empty sum is zero. The final expression includes only candidates completed before $K$, with no benefit in the completion slot. All charged costs remain, including costs of unfinished work. Zero background share can mean capacity gating, budget exhaustion, or no pending work; previously completed artifacts can still contribute benefits.

Exp. CE uses an exponential-utility certainty equivalent, while the conservative rule discounts reuse, qualification, and unit benefit. Their final synthetic net values are 159.89 and 163.16 for this fixed seed. A single trace illustrates delayed amortization and the accounting convention; it does not establish a distribution-wide policy ranking or measure real capability gains from the simulated work.

\section{Finite Model: Specification and Guarantees}
\label{app:model_grid}
\setcounter{figure}{0}
\setcounter{equation}{0}
\setcounter{table}{0}
The finite experiment in Section~\ref{sec:state_eval} compares execution plans under common information, transitions, prices, and return constraints. Exact dynamic programming optimizes this action model; unmodeled arrivals, candidates, and runtime mechanisms remain outside its scope.

\subsection{Scenario Grid and State Semantics}
Each scenario contains one synthetic capability. Its work-size parameter is 30, 60, 120, or 180 s of construction and validation, with completed and durable progress set by the initial state. The primary grid uses a 5 s work quantum. Table~\ref{tab:window_calendars} lists the six ordered calendars; adjacent windows are separated by 20 s, and service must be restored by each window's end. We set the first window's start to zero.

\begin{table}[!htbp]
\centering
\caption{Ordered window durations in the finite model. Adjacent windows have a 20 s gap; the single continuous window has no inter-window gap.}
\label{tab:window_calendars}
\small
\begin{tabular}{@{}ll@{}}
\toprule
Calendar & Window durations (s) \\
\midrule
Front-loaded & 240, 80, 80, 80 \\
Back-loaded & 80, 80, 80, 240 \\
Alternating & 120, 240, 120, 240 \\
Fragmented & 80, 80, 80, 80 \\
Uniform & 160, 160, 160, 160 \\
Continuous & 480 \\
\bottomrule
\end{tabular}
\end{table}

Post-publication benefit rates are 0.1, 0.25, 0.5, 1, or 2 s-eq./s, and the benefit horizon ends 120 or 480 s after the last window. Each combination is evaluated with cold and warm recovery and four initial states: fresh, half-complete with unsaved work, half-complete with saved work while active, and half-complete with saved work in service mode. The distinction between completed and durable progress controls what survives interruption; active states incur return costs even when abandoned.

The Cartesian product has $6\times4\times5\times2\times2\times4=1920$ scenarios. The configuration is frozen and hashed before computation. Means over this complete grid describe its specified mixture of cases; they are not estimates of how frequently those cases occur in a production workload.

\subsection{Transition Costs and Exact Evaluation}
Cold and warm recovery take 69.988 and 0.487 s, using three-run SGLang means. Save and load take 68.075 and 47.748 ms from an independent LoRA microbenchmark. These delays remain exact wall-clock costs even though progress advances in discrete quanta. Execution and recovery cost one s-eq. per second. Warm recovery additionally prices 6.383 GiB of incremental host PSS at 0.01 s-eq. per GiB-second, or 0.06383 s-eq. per second of charged residency.

The measurements come from separate experiments. The memory price is assumed, unmeasured entry and release delays are zero, and resident-device-memory effects on throughput are not fitted. A residency bill depends on retained memory and retention time; recovery latency alone cannot determine net value.

At window $k$, a feasible macro-action must satisfy Equation~\eqref{eq:macro_safe}, including loading, checkpointing when selected, and the return path. Unsaved progress is discarded on a return without checkpointing. Waiting in service mode adds no recovery cost. Publishing at $t_p$ settles the remaining benefit as $\Phi(t_p)=g\pos{H-t_p}$ and still charges the publication action's return. Abandonment earns no capability benefit but pays any required return. Never can therefore have negative value when initially active.

\paragraph{Finite-stage assumptions}
Index stages by $k=0,\ldots,T-1$. The state contains the stage and all information determining transitions and rewards. Each action set $\mathcal A_k(s)$ is finite, nonempty, and return-feasible. Every nonterminal action advances one stage, excluding zero-duration cycles. Publication and abandonment enter an absorbing state with zero subsequent rewards. All rewards are finite, and publication benefit is credited only to the publishing action.

Let $F_k(s,a)$ be the successor and $R_k(s,a)$ the net reward. A feasible policy $\pi$ selects $a_k=\pi_k(s_k)\in\mathcal A_k(s_k)$. Starting from $s_k=s$ and following $s_{j+1}=F_j(s_j,\pi_j(s_j))$, its value satisfies
\begin{equation}
 \begin{aligned}
 V_k^\pi(s)
 &=\sum_{j=k}^{T-1}R_j(s_j,\pi_j(s_j))\\
 &=R_k(s,\pi_k(s))
   +\sum_{j=k+1}^{T-1}R_j(s_j,\pi_j(s_j))\\
 &=R_k(s,\pi_k(s))+V_{k+1}^\pi(s_{k+1}).
 \end{aligned}
 \label{eq:finite_policy_value}
\end{equation}
The empty terminal sum gives $V_T^\pi=0$. For a proposed first action $a$, write $s_a=F_k(s,a)$ and let $\rho$ range over feasible continuation policies from stage $k+1$. Optimizing the first action and then its continuation gives
\begin{equation}
 \begin{aligned}
 V_k^*(s)
 &=\max_{\substack{a\in\mathcal A_k(s),\,\rho}}
   \bigl\{R_k(s,a)+V_{k+1}^{\rho}(s_a)\bigr\}\\
 &=\max_{a\in\mathcal A_k(s)}
   \bigl\{R_k(s,a)+\max_{\rho}V_{k+1}^{\rho}(s_a)\bigr\}\\
 &=\max_{a\in\mathcal A_k(s)}
   \bigl\{R_k(s,a)+V_{k+1}^*(s_a)\bigr\}.
 \end{aligned}
 \label{eq:finite_dp}
\end{equation}
The finite reachable decision tree ensures that the maxima exist. Backward evaluation from $V_T^*=0$ gives the exact feasible optimum. Policy comparisons hold transitions, terminal values, and cost accounting fixed.

\subsection{Policy Improvement and Scoring Error}
Retain the finite-stage assumptions above and the reference $\pi_0$ from Equation~\eqref{eq:rollout}. At each state, let the candidate set $\mathcal C_k(s)\subseteq\mathcal A_k(s)$ contain the feasible reference action $\pi_{0,k}(s)$. Define the exact reference-continuation score and rollout choice by
\begin{equation}
 \begin{aligned}
 Q_k^{\pi_0}(s,a)
 &=R_k(s,a)+V_{k+1}^{\pi_0}(F_k(s,a)),\\
 \pi_{R,k}(s)
 &\in\arg\max_{a\in\mathcal C_k(s)}Q_k^{\pi_0}(s,a).
 \end{aligned}
 \label{eq:finite_rollout}
\end{equation}
Scores use accurate transitions, rewards, and reference values. Ties retain the reference whenever it maximizes the score. The candidate set may omit other feasible actions without weakening the reference-relative guarantee.

\begin{proof}[Proof of Proposition 1]
At stage $T$, both policies have zero continuation value. Assume $V_{k+1}^{\pi_R}(s')\geq V_{k+1}^{\pi_0}(s')$ for every reachable state at the next stage. Fix a stage-$k$ state $s$ and abbreviate $a_R=\pi_{R,k}(s)$, $a_0=\pi_{0,k}(s)$, and $s_R=F_k(s,a_R)$. Then
\begin{equation}
 \begin{aligned}
 V_k^{\pi_R}(s)
 &=R_k(s,a_R)+V_{k+1}^{\pi_R}(s_R)\\
 &\overset{\rm (i)}{\geq}R_k(s,a_R)+V_{k+1}^{\pi_0}(s_R)\\
 &=Q_k^{\pi_0}(s,a_R)\\
 &\overset{\ (ii)}{\geq}Q_k^{\pi_0}(s,a_0)\\
 &=V_k^{\pi_0}(s).
 \end{aligned}
 \label{eq:rollout_proof}
\end{equation}
The outer equalities evaluate each policy. Step (i) uses the induction hypothesis at $s_R$; step (ii) uses score maximization and $a_0\in\mathcal C_k(s)$. Backward induction gives $V_k^{\pi_R}(s)\geq V_k^{\pi_0}(s)$ at every reachable stage and state.
\end{proof}
A gap to $V_k^*$ can remain because rollout scores use the reference continuation rather than an optimal continuation.

\paragraph{Bounded scoring error}
Now let $\widehat Q_k$ estimate the same true scores. Assume a uniform bound $|\widehat Q_k(s,a)-Q_k^{\pi_0}(s,a)|\leq\epsilon$ with $\epsilon\geq0$, covering every candidate and reference action at all states reachable under the considered policies. At a fixed $(k,s)$, suppress those indices and set $e_a=\widehat Q(a)-Q^{\pi_0}(a)$. Define $A(a)=Q^{\pi_0}(a)-Q^{\pi_0}(a_0)$ and $\widehat A(a)=\widehat Q(a)-\widehat Q(a_0)$. The two score errors imply
\begin{equation}
 \begin{aligned}
 A(a)
 &=[\widehat Q(a)-e_a]-[\widehat Q(a_0)-e_{a_0}]\\
 &=\widehat A(a)-e_a+e_{a_0}\\
 &\geq\widehat A(a)-|e_a|-|e_{a_0}|\\
 &\geq\widehat A(a)-2\epsilon.
 \end{aligned}
 \label{eq:rollout_error_margin}
\end{equation}
Define a guarded policy $\pi_G$ that selects a replacement $a\in\mathcal C_k(s)$ only if $\widehat A(a)>2\epsilon$ and otherwise retains $a_0$. An accepted replacement has $A(a)>0$ by Equation~\eqref{eq:rollout_error_margin}; retaining $a_0$ gives $A(a_0)=0$. Hence every decision has a true reference-relative score at least zero.

\begin{proof}[Policy guarantee with estimated scores]
The terminal values again coincide. Assuming improvement at every next-stage state and writing $a_G=\pi_{G,k}(s)$, $s_G=F_k(s,a_G)$, we obtain
\begin{equation}
 \begin{aligned}
 V_k^{\pi_G}(s)
 &=R_k(s,a_G)+V_{k+1}^{\pi_G}(s_G)\\
 &\geq R_k(s,a_G)+V_{k+1}^{\pi_0}(s_G)\\
 &=Q_k^{\pi_0}(s,a_G)\\
 &=V_k^{\pi_0}(s)+A(a_G)\\
 &\geq V_k^{\pi_0}(s).
 \end{aligned}
 \label{eq:guarded_rollout_proof}
\end{equation}
Backward induction therefore extends the local score condition to the value of the whole guarded policy. This argument requires the error bound wherever the policy can visit, including states reached after earlier replacements.
\end{proof}
The experiment uses accurate scores; bounds on online estimation errors require separate validation.

\subsection{Paired Comparisons and Resolution}
The informative policy difference is $V^{\pi_R}(s)-V^{\pi_0}(s)$ under identical state and calendar inputs. Separate gaps to $V^*(s)$ reveal whether both policies are suboptimal. Table~\ref{tab:resolution} reports the full grid and the finer-quantum check. All 91 strict improvements in the primary grid involve cold recovery, while all 960 warm-recovery cases tie with Costed Window. Rollout matches DP in 1907 cases; the other 13 have a maximum gap of 124.836 s-eq. A small mean gap therefore does not imply uniform near-optimality.

\begin{table}[!htbp]
\centering
\caption{Rollout versus Costed Window, with no regressions. Mean DP gaps use s-eq.; 1 s cases start fresh.}
\label{tab:resolution}
\small
\setlength{\tabcolsep}{3pt}
\begin{tabular}{@{}lrrr@{}}
\toprule
Evaluation & Wins / ties & \shortstack{Rollout\\gap} & \shortstack{Costed\\Window gap} \\
\midrule
Full grid, 5 s & 91 / 1829 & 0.246 & 5.369 \\
Fresh only, 1 s & 27 / 453 & 0 & 5.869 \\
\bottomrule
\end{tabular}
\end{table}

The 1 s experiment refines only the 480 fresh-candidate cases, holding their other parameters fixed. Its zero rollout gap establishes agreement with DP on that finer finite action set. It does not establish an optimum in continuous time, cover partially completed initial states, or prove that rollout is optimal for other calendars.

\clearpage
\end{document}